\documentclass[aps, prd, showkeys ]{revtex4}
\usepackage[small]{caption}
\usepackage{amsmath,amsfonts,amscd,amssymb,epsf, epsfig,mathrsfs}\usepackage{graphicx}

\newcommand{\pa}{\partial}

\begin{document}
 \title{Finite temperature Casimir effect of massive fermionic fields in the presence of compact dimensions}
 \author{F. S. Khoo}\email{ Fech.Scen@nottingham.edu.my}
\address{Department of Applied Mathematics, Faculty of Engineering, University of Nottingham Malaysia Campus, Jalan Broga, 43500, Semenyih, Selangor Darul Ehsan, Malysia.}
\author{L. P. Teo}\email{ LeePeng.Teo@nottingham.edu.my}
\address{Department of Applied Mathematics, Faculty of Engineering, University of Nottingham Malaysia Campus, Jalan Broga, 43500, Semenyih, Selangor Darul Ehsan, Malysia.}

\begin{abstract}
We consider the finite temperature Casimir effect   of a massive fermionic field confined between two parallel plates, with MIT bag boundary conditions on the plates. The background spacetime is $M^{p+1}\times T^q$ which has $q$ dimensions compactified to a torus. On the compact dimensions, the field is assumed to satisfy   periodicity boundary conditions with arbitrary phases. Both the high temperature and the low temperature expansions of the Casimir free energy and the force are derived explicitly. It is found that the Casimir force acting on the plates is always attractive at any temperature regardless of the boundary conditions assumed on the compact torus. The asymptotic limits of the Casimir force in the small plate separation limit are also obtained.

\end{abstract}

\maketitle

\section{Introduction}

Casimir effect is a manifestation of the zero-point energy of a quantum field \cite{9,11,10}. Since spacetimes with extra dimensions are fundamental in most of the theories of high energy physics,   there have been some intensive activities in investigating the Casimir effect in spacetime with extra compactified dimensions \cite{12,13,14,15,16,17,18,19,20,21,22,23,32,24,25,1,26,41,7,27,6,28,29,30,31,33}. The case of scalar fields with Dirichlet or Neumann boundary conditions or more general Robin conditions, and the case of electromagnetic field with perfectly conducting or infinitely permeable boundary conditions have been  studied at either   zero or finite temperature, for different extra-dimensional spacetimes such as the Kaluza-Klein spacetime and the Randall-Sundrum spacetime. One of the common findings of these works is that on a pair of parallel plates with the same boundary conditions, the Casimir force is always attractive at any temperature. For fermionic fields, so far only the zero temperature effect has been investigated in \cite{41,7,27,32} for extra-dimensional spacetimes with toroidal extra dimensions. It was found that for a massive fermionic field with MIT bag boundary conditions on a pair of parallel plates, the zero temperature Casimir force is attractive regardless of the boundary conditions imposed on the compactified dimensions. The temperature correction to the Casimir force of a fermionic field is different from that of a bosonic field. Therefore, it would be interesting to investigate whether the Casimir force of a fermionic field would stay attractive at any temperature. The purpose of this work is to answer this question.

The zero temperature Casimir effect of fermionic fields in (3+1)-dimensional Minkowski spacetime has been investigated in \cite{34,38,40} for the massless case and in \cite{35,43} for the massive case. The zero temperature effect in Minkowski spacetime with arbitrary number of dimensions was discussed  in \cite{36,37}. The finite temperature correction to the fermionic Casimir effect is less considered. For   massless fermions    in (3+1)-dimensional Minkowski spacetime, it  was discussed in \cite{42,39}. In this article, we consider the finite temperature Casimir effect of a massive fermionic field in a spacetime of the form $M^{p+1}\times T^q$, where $M^{p+1}$ is the $(p+1)$-dimensional Minkowski spacetime and $T^q$ is a compact $q$-dimensional torus. This will cover the massless case by taking the limit $m\rightarrow \infty$ and the case of a Minskowki spacetime with arbitrary dimensions by setting $q=0$ or by letting the radii of the torus go to infinity.

In this paper, we use units with $\hbar=c=k_B=1$.

\section{The Casimir free energy}

As in \cite{7}, we consider a quantum fermionic field $\psi$ on a  ($D+1$)-dimensional   spacetime    with $q$ compact dimensions of the form  $M^{p+1}\times T^q$, where $M^{p+1}$ is the $(p+1)$-dimensional Minkowski spacetime, and $T^q=(S^1)^q$ is a $q$ torus.  The background metric is the flat metric $$ds^2=g_{\mu\nu}dx^{\mu}dx^{\nu}=dt^2-\sum_{j=1}^D(dx^j)^2.$$Here $t=x^0$,  $x^j\in \mathbb{R}$ for $j=0,1,\ldots,p$ and $0\leq x^j\leq L_j$ for $j=p+1,\ldots,D$. The field $\psi$ has $N_D$ components, where $N_D$ is given by $2^{\frac{D+1}{2}}$ if $D$ is odd and $2^{\frac{D}{2}}$ if $D$ is even.
On the compact dimensions,
  the field $\psi$ is assumed to satisfy the general periodicity conditions
\begin{equation}\label{eq6_22_1}
\psi(t, \boldsymbol{x} +L_j\boldsymbol{e}_j)=e^{2\pi i\alpha_j}\psi(t,\boldsymbol{x}),
\end{equation}
where $\boldsymbol{x}=(x^1,\ldots,x^D)$, $\boldsymbol{e}_j$ is the unit vector in the $x^j$ direction, and $0\leq \alpha_j<1, j=p+1,\ldots,D,$ are the constant phases. The cases $\alpha_j=0$ for all $j=p+1,\ldots,D,$ and $\alpha_j=1/2$ for all $j=p+1,\ldots,D,$ correspond respectively  to untwisted    and twisted fields.

In this paper, we consider the finite temperature Casimir effect of the fermionic field $\psi$ when it is confined between two parallel plates placed at $x^1=0$ and $x^1=a$.
The equation of motion of the field $\psi$ is the Dirac equation:
\begin{equation}
 i \gamma^\mu\partial_\mu\psi - m\psi = 0.
\end{equation}On the plates $x^1=0$ and $x^1=a$, the field $\psi$ satisfies the MIT bag boundary conditions:
\begin{equation}\label{eq6_22_2}
\left.(1+ i \gamma^{\mu}n_{\mu})\psi\right|_{x^1=0\;\text{and} \;x^1=a}\psi = 0.
\end{equation}
Here $\gamma^{\mu}$ are the Dirac matrices, and $n_{\mu}$ is a unit outward normal vector to the boundaries.

As in \cite{7}, using the chiral representation of the Dirac matrices:
\begin{equation*}
\gamma^0=\begin{pmatrix} 1 & 0\\ 0 &-1\end{pmatrix},\hspace{1cm}\gamma^{j}=\begin{pmatrix} 0 &\sigma_j\\-\sigma_j^{+} & 0\end{pmatrix},\quad j=1,\ldots,D,
\end{equation*}
with  $\sigma_j\sigma_l^{+}+\sigma_l\sigma_j^{+}=2\delta_{jl}$, the positive-frequency and the negative-frequency solutions of the Dirac equation can be written respectively as
\begin{equation}	
\psi^{(+)} =  e^{- i\omega t}\begin{pmatrix}
\varphi^{(+)}\\
- i \boldsymbol{\sigma}^{+}\boldsymbol{\cdot}\boldsymbol{\nabla}\varphi^{(+)}/(\omega+m) \end{pmatrix}
\quad\text{and}
\quad\psi^{(-)} =  e^{ i\omega t}\begin{pmatrix}
 i \boldsymbol{\sigma}\cdot\boldsymbol{\nabla}\varphi^{(-)}/(\omega+m)\\
\varphi^{(-)} \end{pmatrix}.
\end{equation}
 Here  $\boldsymbol{\sigma} = (\sigma^1,\dotsc, \sigma^D)$, the spinors $\varphi^{(+)}$ and $\varphi^{(-)}$ are given   by
\begin{equation}
\varphi^{(\pm)} =
\left(\varphi_+^{(\pm)}e^{ i  k_{1}x^{ 1}} + \varphi_-^{(\pm)}e^{- i  k_{ 1}x^{ 1}}\right)\exp\left(\pm i\sum_{j=2}^Dk_jx^j\right),
\end{equation}and
$\displaystyle\omega^2=\sum_{j=1}^Dk_j^2+m^2.$
The boundary conditions on the compact dimensions \eqref{eq6_22_1} imply that
\begin{equation*}
k_j=\frac{2\pi (n_j+\alpha_j)}{L_j}\quad\text{for}\,j=p+1,\ldots,D,
\end{equation*}where $n_j$ are integers. On the uncompactified directions $x^2,\ldots,x^p$, there are no boundary conditions and hence $k_2,\ldots, k_p\in \mathbb{R}$.
The boundary condition \eqref{eq6_22_2} on the plate located at $x^1=0$ implies that
\begin{equation}\label{eq6_23_3}\begin{split}
\varphi_+^{(\pm)}=&-\frac{m(\omega+m)+k_1^2\mp k_1\sigma_1\boldsymbol{\sigma}_{\parallel}^+\cdot\boldsymbol{k}_{\parallel}}{(m-ik_1)(\omega+m)}\varphi_-^{(\pm)},
\end{split}\end{equation}whereas the boundary conditions on the plate located at $x^1=a$ implies
\begin{equation}\label{eq6_23_4}\begin{split}
\varphi_+^{(\pm)}=&-\frac{m(\omega+m)+k_1^2\mp k_1\sigma_1\boldsymbol{\sigma}_{\parallel}^+\cdot\boldsymbol{k}_{\parallel}}{(m+ik_1)(\omega+m)}e^{-2ik_1a}\varphi_-^{(\pm)}.
\end{split}\end{equation}Here $\boldsymbol{\sigma}_{\parallel}=(\sigma_2,\ldots,\sigma_D)$ and $\boldsymbol{k}_{\parallel}=(k_2,\ldots,k_D)$.
Comparing \eqref{eq6_23_3} and \eqref{eq6_23_4}, we find that in order to have nontrivial solutions for $(\varphi_+^{(+)},\varphi_-^{(+)})$ and $(\varphi_+^{(-)},\varphi_-^{(-)})$, one requires
  $k_1$ to satisfy a transcendental equation
\begin{equation}\label{eq6_22_3}
 F(z):=m\sin az+z\cos  az=0.
\end{equation}
Therefore, the   eigenfrequencies of the field $\psi$ are given by
\begin{gather*}\omega=\sqrt{k_1^2+\sum_{j=2}^pk_j^2+\lambda_{\boldsymbol{n}}^2+m^2},\hspace{1cm}
\lambda_{\boldsymbol{n}}^2=\sum_{j=p+1}^D\left[\frac{2\pi(n_j+\alpha_j)}{L_j}\right]^2,\quad\boldsymbol{n}=(n_{p+1},\ldots,n_D),\end{gather*} where
$k_1$ are positive solutions of \eqref{eq6_22_3}, $k_j\in \mathbb{R}$ for $j=2,\ldots,p$, and $n_j, j=p+1,\ldots, D$, are integers. Each of these $\omega$ appears with multiplicity $N_D$.

\begin{figure}[h]
\epsfxsize=0.5\linewidth \epsffile{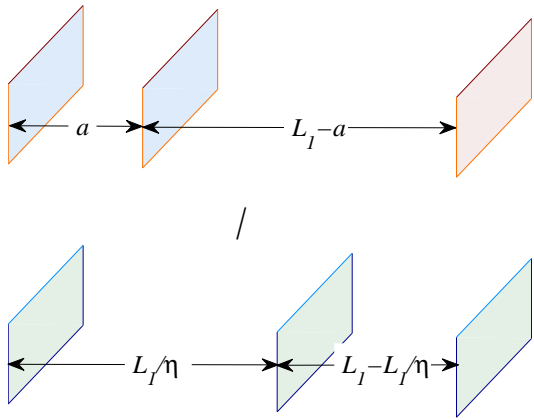} \caption{\label{f1} The piston regularization scheme. }\end{figure}

To regularize the Casimir free energy of the parallel plate system, we take the piston approach, where the Casimir free energy is given by \cite{8}:
\begin{equation}\label{eq6_22_6}
E_{\text{Cas}}^{\parallel}=\lim_{L_1\rightarrow\infty}\Bigl(E_{\text{Cas}} (a)+E_{\text{Cas}} (L_1-a)-E_{\text{Cas}} \left(L_1/\eta\right)-E_{\text{Cas}} \left(L_1\left[1-1/\eta\right]\right)\Bigr).
\end{equation}See Figure \ref{f1}. Here $E_{\text{Cas}} (a)$ is the Casimir free energy between the parallel plates which are separated by a distance $a$, and $\eta$ is a constant greater than 1.
Using Matsubara imaginary time formalism, for a fermionic system in thermal equilibrium at temperature $T$, the partition function is given by
\begin{equation*}
\mathscr{Z}=\int D\psi D\bar{\psi}\exp\left(\int_0^{\beta}d\tau\int d^D\boldsymbol{x} \mathcal{L}\right),
\end{equation*}where $$\mathcal{L}=\frac{i}{2}\left(\bar{\psi}\gamma^{\mu}\pa_{\mu}\psi-\pa_{\mu}\bar{\psi}\gamma^{\mu} \psi\right)-m\bar{\psi}\psi$$ is the Dirac Lagrangian density. The time direction $t$ is rotated to the imaginary axis, i.e., $t\mapsto i\tau$, and  compactified to a circle of radius $\beta=1/T$. The field $\psi$ satisfies the  antiperiodic boundary conditions on the imaginary time direction: $$\psi(x,\tau+\beta)=-\psi(x,\tau).$$
The finite temperature
Casimir free energy is then given by
\begin{align*}
E_{\text{Cas}}=-\frac{1}{\beta}\ln \mathscr{Z}=-T\ln\mathscr{Z}.
\end{align*}
 Using zeta function techniques \cite{8,45,44}, we have
\begin{equation}
E_{\text{Cas}}  = \frac{T}{2}\left(\zeta'_T(0) + [\log \mu^2]\zeta_T(0)\right),
\end{equation}
where $\mu$ is a normalization constant with the dimension of mass, and $\zeta_T(s)$ is the zeta function
\begin{equation*}
\zeta_T(s)=\sum_{\omega\;\text{eigenfrequencies}}\sum_{l=-\infty}^{\infty}\left(\omega^2+\xi_l^2\right)^{-s}.
\end{equation*}Here $\displaystyle\xi_l= 2\pi\left(l+\frac{1}{2}\right) T$  are the Matsubara frequencies.
For the Casimir free energy between the parallel plates which are separated by a distance $a$,
 \begin{equation*}\begin{split}
\zeta_T(s;a)=&\frac{AN_D}{(2\pi)^{p-1}}\sum_{l=-\infty}^{\infty}\sum_{\boldsymbol{n}\in\mathbb{Z}^q}\sum_{k_1}\int_{\mathbb{R}^{p-1}}
\left(k_1^2+\sum_{j=2}^pk_j^2+\lambda_{\boldsymbol{n}}^2+\xi_l^2+m^2\right)^{-s}dk_2\ldots dk_p\\
=&\frac{AN_D\Gamma\left(s-\frac{p-1}{2}\right)}{2^{p-1}\pi^{\frac{p-1}{2}}\Gamma\left(s\right)}\sum_{l=-\infty}^{\infty}\sum_{\boldsymbol{n}\in\mathbb{Z}^q}\sum_{k_1}
\left(k_1^2+ m_{\boldsymbol{n},l}^2\right)^{-s+\frac{p-1}{2}},
\end{split}\end{equation*}where $A$ is the area of the plates and $m_{\boldsymbol{n},l}^2=\lambda_{\boldsymbol{n}}^2+\xi_l^2+m^2$.
In order to perform the summation over $k_1$, which are positive solutions of the equation $F(z)=0$, we need to use   the generalized Abel-Plana summation formula, which states that \cite{4,5,6}: If $f_0(z)$, $f_+(z)$ and $f_-(z)$ are functions such that
\begin{equation}\label{eq6_24_2}\begin{split}
\lim_{Y\rightarrow \infty} \int_{0}^{\infty}\Bigl\{ f_0(x\pm iY)- f_{\pm}(x\pm iY)\Bigr\}dx=0,\hspace{1cm}\lim_{X\rightarrow \infty} \int_{0}^{\infty}\Bigl\{ f_0(X\pm iy)- f_{\pm}(X\pm iy)\Bigr\}dy=0,
\end{split}
\end{equation} then
\begin{equation}\label{eq6_24_1}
\begin{split}
&\sum_{  \text{Re}\; z\geq 0} w_0(z) \text{Res}_{z}f_0(z) -\sum_{\substack{ \text{Re}\; z\geq 0\\ \text{Im}\;z\geq 0}} w_+(z) \text{Res}_{z}f_+(z)
 -\sum_{\substack{ \text{Re}\; z\geq 0\\ \text{Im}\;z\leq 0}} w_-(z)  \text{Res}_{z}f_-(z) \\=& - \frac{1}{2\pi }\int_0^{\infty} \Bigl\{
 f_0(iy)-f_+(iy)\Bigr\}dy- \frac{1}{2\pi }\int_0^{\infty} \Bigl\{
 f_0(-iy)-f_-(-iy)\Bigr\}dy-\frac{1}{2\pi i}\int_0^{\infty} \Bigl\{f_+(x)-f_-(x)\Bigr\}dx.
\end{split}
\end{equation} Here $w_0(z), w_1(z)$ and $w_2(z)$ are weight functions defined by
\begin{equation*}
\begin{split}
w_0(z)&=\begin{cases}1, \hspace{0.5cm}&\text{if}\quad\text{Re}\; z> 0, \\
 1/2, &\text{if}\quad\text{Re}\; z= 0\end{cases},\hspace{1cm} w_+(z) =\begin{cases} 1, \hspace{0.5cm}&\text{if}\quad\text{Re}\;z>0\;\text{and}\;\text{Im}\;z>0,\\
   1/2, &\text{if} \quad \text{Re}\;z=0\;\text{or}\;\text{Im}\;z=0,\\
   1/4, &\text{if}\quad z=0,\end{cases},
\end{split}\end{equation*}and $w_-(z)=w_+(\bar{z})$.
Notice that
$$F(z)=\frac{z-im}{2}e^{iaz}+\frac{z+im}{2}e^{-iaz}.$$
Taking
\begin{equation*}\begin{split}
f_0(z)=&\frac{AN_D\Gamma\left(s-\frac{p-1}{2}\right)}{2^{p-1}\pi^{\frac{p-1}{2}}\Gamma\left(s\right)}\sum_{l=-\infty}^{\infty}\sum_{\boldsymbol{n}\in\mathbb{Z}^q}
\left(z^2+ m_{\boldsymbol{n},l}^2\right)^{-s+\frac{p-1}{2}}\frac{d}{dz}\ln F(z),\\
f_{\pm}(z)=&\frac{AN_D\Gamma\left(s-\frac{p-1}{2}\right)}{2^{p-1}\pi^{\frac{p-1}{2}}\Gamma\left(s\right)}\sum_{l=-\infty}^{\infty}\sum_{\boldsymbol{n}\in\mathbb{Z}^q}
\left(z^2+ m_{\boldsymbol{n},l}^2\right)^{-s+\frac{p-1}{2}}\frac{d}{dz}\ln \left(\frac{z\pm im}{2}e^{\mp ia z}\right),
\end{split}\end{equation*}we find that the conditions \eqref{eq6_24_2} are satisfied. Since the poles of $f_0(z)$ are the zeros of the function $F(z)$, and the function $f_+(z)$ (resp. $f_-(z)$) does not have poles on the upper (resp. lower) half plane, the  left hand side of \eqref{eq6_24_1} gives
\begin{equation}\label{eq6_22_4}
\zeta_T(s;a)+\frac{1}{2}\frac{AN_D\Gamma\left(s-\frac{p-1}{2}\right)}{2^{p-1}\pi^{\frac{p-1}{2}}\Gamma\left(s\right)}
\sum_{l=-\infty}^{\infty}\sum_{\boldsymbol{n}\in\mathbb{Z}^q}
 m_{\boldsymbol{n},l}^{-2s+p-1},
\end{equation}where the second term comes from the pole of $f_0(z)$ at $z=0$. This term is independent of $a$. On the other hand,
\begin{equation*}
\begin{split}
f_0(\pm iy)-f_{\pm}(\pm iy)=&\frac{AN_D\Gamma\left(s-\frac{p-1}{2}\right)}{2^{p-1}\pi^{\frac{p-1}{2}}\Gamma\left(s\right)}\sum_{l=-\infty}^{\infty}\sum_{\boldsymbol{n}\in\mathbb{Z}^q}
\frac{d}{dy}\ln\left(1+\frac{y-m}{y+m}e^{-2ay}\right)\\&\times\left\{\begin{aligned}&\mp i\left(m_{\boldsymbol{n},l}^2-y^2\right)^{-s+\frac{p-1}{2}},\quad & y<m_{\boldsymbol{n},l}\\
&\mp i\left(y^2-m_{\boldsymbol{n},l}^2 \right)^{-s+\frac{p-1}{2}}e^{\mp i\pi\left(s-\frac{p-1}{2}\right)},\quad &y>m_{\boldsymbol{n},l}\end{aligned}\right.,
\end{split}
\end{equation*}
\begin{equation*}
\begin{split}
f_+(x)-f_-(x)=&\frac{AN_D\Gamma\left(s-\frac{p-1}{2}\right)}{2^{p-1}\pi^{\frac{p-1}{2}}\Gamma\left(s\right)}\sum_{l=-\infty}^{\infty}\sum_{\boldsymbol{n}\in\mathbb{Z}^q}
\left(x^2+ m_{\boldsymbol{n},l}^2\right)^{-s+\frac{p-1}{2}}\left(-\frac{2im}{x^2+m^2}-2ia\right).
\end{split}
\end{equation*}Therefore, the right hand side of \eqref{eq6_24_1} gives
\begin{equation}\label{eq6_22_5}
\begin{split}
&\frac{\sin\pi\left(s-\frac{p-1}{2}\right)}{\pi}\frac{AN_D\Gamma\left(s-\frac{p-1}{2}\right)}{2^{p-1}\pi^{\frac{p-1}{2}}\Gamma\left(s\right)}
\sum_{l=-\infty}^{\infty}\sum_{\boldsymbol{n}\in\mathbb{Z}^q}\int_{m_{\boldsymbol{n},l}}^{\infty}\left(y^2-m_{\boldsymbol{n},l}^2 \right)^{-s+\frac{p-1}{2}}
\frac{d}{dy}\ln\left(1+\frac{y-m}{y+m}e^{-2ay}\right)dy\\
&+ \frac{AN_D\Gamma\left(s-\frac{p-1}{2}\right)}{2^{p-1}\pi^{\frac{p+1}{2}}\Gamma\left(s\right)}\sum_{l=-\infty}^{\infty}\sum_{\boldsymbol{n}\in\mathbb{Z}^q}
\int_0^{\infty}\left(x^2+ m_{\boldsymbol{n},l}^2\right)^{-s+\frac{p-1}{2}}\left( \frac{m}{x^2+m^2}+a\right)dx.
\end{split}
\end{equation}Notice that the second term depends on $a$ linearly. From \eqref{eq6_22_4} and \eqref{eq6_22_5}, one finds that
\begin{equation*}
\zeta_T(s;a)=\zeta_{T;0}(s)+a\zeta_{T;1}(s)+ \frac{AN_D}{2^{p-1}\pi^{\frac{p-1}{2}}\Gamma\left( \frac{p+1}{2}-s\right)\Gamma\left(s\right)}
\sum_{l=-\infty}^{\infty}\sum_{\boldsymbol{n}\in\mathbb{Z}^q}\int_{m_{\boldsymbol{n},l}}^{\infty}\left(y^2-m_{\boldsymbol{n},l}^2 \right)^{-s+\frac{p-1}{2}}
\frac{d}{dy}\ln\left(1+\frac{y-m}{y+m}e^{-2ay}\right)dy.
\end{equation*}Here $\zeta_{T;0}(s)$ and $\zeta_{T;1}(s)$ do not depend on $a$. From this, we have
\begin{equation*}
\begin{split}
\zeta_T(0;a)=&\zeta_{T;0}(0)+a\zeta_{T;1}(0),\\
\zeta_T'(0;a)=&\zeta_{T;0}'(0)+a\zeta_{T;1}'(0)+ \frac{AN_D}{2^{p-1}\pi^{\frac{p-1}{2}} \Gamma\left( \frac{p+1}{2} \right)}
\sum_{l=-\infty}^{\infty}\sum_{\boldsymbol{n}\in\mathbb{Z}^q}\int_{m_{\boldsymbol{n},l}}^{\infty}\left(y^2-m_{\boldsymbol{n},l}^2 \right)^{ \frac{p-1}{2}}
\frac{d}{dy}\ln\left(1+\frac{y-m}{y+m}e^{-2ay}\right)dy,
\end{split}
\end{equation*}  and thus
\begin{equation*}
E_{\text{Cas}}(a)=\mathcal{E}_0+a\mathcal{E}_1-\frac{ATN_D(p-1)}{2^{p}\pi^{\frac{p-1}{2}}\Gamma\left( \frac{p+1}{2} \right) }
\sum_{l=-\infty}^{\infty}\sum_{\boldsymbol{n}\in\mathbb{Z}^q}\int_{m_{\boldsymbol{n},l}}^{\infty}y\left(y^2-m_{\boldsymbol{n},l}^2 \right)^{ \frac{p-3}{2}}
 \ln\left(1+\frac{y-m}{y+m}e^{-2ay}\right)dy,
\end{equation*}where $\mathcal{E}_0$ and $\mathcal{E}_1$ are independent of $a$. After regularization using the piston scheme \eqref{eq6_22_6}, the terms $\mathcal{E}_0$ and $\mathcal{E}_1$ are canceled, and we find that the Casimir free energy between the parallel plates is given by
\begin{equation}\label{eq6_23_1}
E_{\text{Cas}}^{\parallel}= -\frac{ATN_D }{2^{p-1}\pi^{\frac{p-1}{2}} \Gamma\left( \frac{p-1}{2} \right)}
\sum_{l=-\infty}^{\infty}\sum_{\boldsymbol{n}\in\mathbb{Z}^q}\int_{m_{\boldsymbol{n},l}}^{\infty}y\left(y^2-m_{\boldsymbol{n},l}^2 \right)^{ \frac{p-3}{2}}
 \ln\left(1+\frac{y-m}{y+m}e^{-2ay}\right)dy.
\end{equation}Since $m_{\boldsymbol{n},l}>m$, the Casimir free energy is always negative.
The expression \eqref{eq6_23_1} also shows that in the high temperature (i.e., $aT\gg 1$) limit, the Casimir free energy is dominated by the terms with Matsubara frequency zero, i.e., the term
\begin{gather}
E_{\text{Cas}}^{\text{cl}}= -\frac{ATN_D }{2^{p-1}\pi^{\frac{p-1}{2}} \Gamma\left( \frac{p-1}{2} \right)}
 \sum_{\boldsymbol{n}\in\mathbb{Z}^q}\int_{m_{\boldsymbol{n}}}^{\infty}y\left(y^2-m_{\boldsymbol{n}}^2 \right)^{ \frac{p-3}{2}}
 \ln\left(1+\frac{y-m}{y+m}e^{-2ay}\right)dy,\\
 m_{\boldsymbol{n}}^2=\lambda_{\boldsymbol{n}}^2+m^2,\nonumber
\end{gather}which is linear in $T$.  This term is called the classical term since it is independent of the Planck constant $\hbar$.

To study the low temperature (i.e., $aT\ll 1$) behavior, we need an alternative expression  for the Casimir free energy which we derive in Appendix \ref{a1}. We find that
\begin{equation*}
E_{\text{Cas}}^{\parallel}=E_{\text{Cas}}^{\parallel,T=0}+\Delta_TE_{\text{Cas}}^{\parallel},
\end{equation*}where $E_{\text{Cas}}^{\parallel,T=0}$ is the zero temperature Casimir energy given by
\begin{equation}\label{eq6_23_8}
E_{\text{Cas}}^{\parallel,T=0}= -\frac{AN_D }{2^{p}\pi^{\frac{p}{2}} \Gamma\left( \frac{p}{2} \right)}
\sum_{\boldsymbol{n}\in\mathbb{Z}^q}\int_{m_{\boldsymbol{n}}}^{\infty}y\left(y^2- m_{\boldsymbol{n}}^2 \right)^{\frac{p-2}{2}}\ln\left(1+\frac{y-m}{y+m}e^{-2ay}\right)dy,
\end{equation}and $\Delta_TE_{\text{Cas}}^{\parallel}$ is the temperature correction given by
\begin{equation}\label{eq6_23_9}\begin{split}
\Delta_TE_{\text{Cas}}^{\parallel}= &-\frac{AN_D }{2^{p-2}\pi^{\frac{p-1}{2}}}
\sum_{\boldsymbol{n}\in\mathbb{Z}^q}\sum_{l=1}^{\infty}(-1)^l
\Biggl\{\frac{a}{4\pi } \left(\frac{2m_{\boldsymbol{n}}T}{l}\right)^{\frac{p+1}{2}}K_{\frac{p+1}{2}}\left(\frac{lm_{\boldsymbol{n}}}{T}\right)
\\&\hspace{3cm}+\frac{1}{2\pi^{\frac{3}{2}} }\int_0^{\infty} \frac{m}{y^2+ m^2} \left(\frac{2T\sqrt{y^2+m_{\boldsymbol{n}}^2}}{l}\right)^{\frac{p}{2}}K_{\frac{p}{2}}\left(\frac{l\sqrt{y^2+m_{\boldsymbol{n}}^2}}{T}\right)dy\\
&\hspace{3cm}-\frac{1}{2\sqrt{\pi} }\sum_{k_1>0\,:\,F(k_1)=0}  \left(\frac{2T\sqrt{k_1^2+m_{\boldsymbol{n}}^2}}{l}\right)^{\frac{p}{2}}K_{\frac{p}{2}}\left(\frac{l\sqrt{k_1^2+m_{\boldsymbol{n}}^2}}{T}\right)\Biggr\}.
\end{split}\end{equation}
In the case $m_{\boldsymbol{n}}=0$, which happens if and only if the field is massless and untwisted, and $\boldsymbol{n}=0$, the term
$$\left(\frac{2m_{\boldsymbol{n}}T}{l}\right)^{\frac{p+1}{2}}K_{\frac{p+1}{2}}\left(\frac{lm_{\boldsymbol{n}}}{T}\right)$$ is understood as
\begin{equation}\label{eq6_28_4}
\lim_{m_{\boldsymbol{n}}\rightarrow0}\left(\frac{2m_{\boldsymbol{n}}T}{l}\right)^{\frac{p+1}{2}}K_{\frac{p+1}{2}}\left(\frac{lm_{\boldsymbol{n}}}{T}\right)
=\frac{\Gamma\left(\frac{p+1}{2}\right)}{2}\left(\frac{2T}{l}\right)^{p+1}.
\end{equation}Therefore, for an untwisted massless fermionic field, the temperature correction contains a term of order $T^{p+1}$.

\section{The Casimir force and its asymptotic behavior at small separation}
The Casimir force acting on the plates induced by the vacuum fluctuation of the field is given by
\begin{equation*}
F_{\text{Cas}}^{\parallel}=-\frac{\pa E_{\text{Cas}}^{\parallel}}{\pa a}.
\end{equation*}Using \eqref{eq6_23_1}, we have
\begin{equation}\label{eq6_23_10}
F_{\text{Cas}}^{\parallel}= -\frac{ATN_D }{2^{p-2}\pi^{\frac{p-1}{2}} \Gamma\left( \frac{p-1}{2} \right)}
\sum_{l=-\infty}^{\infty}\sum_{\boldsymbol{n}\in\mathbb{Z}^q}\int_{m_{\boldsymbol{n},l}}^{\infty}\frac{y^2\left(y^2-m_{\boldsymbol{n},l}^2 \right)^{ \frac{p-3}{2}}
}{\frac{y+m}{y-m}e^{2ay}+1}dy.
\end{equation} On the other hand, using \eqref{eq6_23_8} and \eqref{eq6_23_9}, we find that the Casimir force can be written as a sum of the zero temperature term and the temperature correction term, where the zero temperature term is
\begin{equation}\label{eq6_23_11}
F_{\text{Cas}}^{\parallel,T=0}= -\frac{AN_D }{2^{p-1}\pi^{\frac{p}{2}} \Gamma\left( \frac{p}{2} \right)}
\sum_{\boldsymbol{n}\in\mathbb{Z}^q}\int_{m_{\boldsymbol{n}}}^{\infty}\frac{y^2\left(y^2- m_{\boldsymbol{n}}^2 \right)^{\frac{p-2}{2}}}{\frac{y+m}{y-m}e^{2ay}+1}dy,
\end{equation}and the thermal correction term is
\begin{equation}\label{eq6_23_12}\begin{split}
\Delta_TF_{\text{Cas}}^{\parallel}= & \frac{AN_D }{2^{p-2}\pi^{\frac{p-1}{2}}}
\sum_{\boldsymbol{n}\in\mathbb{Z}^q}\sum_{l=1}^{\infty}(-1)^l
\Biggl\{\frac{1}{4\pi } \left(\frac{2m_{\boldsymbol{n}}T}{l}\right)^{\frac{p+1}{2}}K_{\frac{p+1}{2}}\left(\frac{lm_{\boldsymbol{n}}}{T}\right)
\\&\hspace{3cm}-\frac{1}{\sqrt{\pi} }\sum_{k_1>0\,:\,F(k_1)=0}  \frac{k_1^2}{a+\frac{m}{m^2+k_1^2}} \left(\frac{2T\sqrt{k_1^2+m_{\boldsymbol{n}}^2}}{l}\right)^{\frac{p-2}{2}}K_{\frac{p-2}{2}}\left(\frac{l\sqrt{k_1^2+m_{\boldsymbol{n}}^2}}{T}\right)\Biggr\}.
\end{split}\end{equation}
Here we have used
\begin{equation*}
\frac{\pa k_1}{\pa a}=-\left.\frac{\frac{\pa F}{\pa a}}{\frac{\pa F}{\pa z}}\right|_{z=k_1}=-\frac{k_1}{a+\frac{m}{m^2+k_1^2}},
\end{equation*}which follows from $F(k_1)=0$.

Since $m_{\boldsymbol{n},l}>m$, the expression \eqref{eq6_23_10} shows manifestly that the Casimir force is always attractive (negative) at any temperature and for any mass $m$.

In the massless case,  the integrals in \eqref{eq6_23_10} and \eqref{eq6_23_11} can be evaluated explicitly, which gives
\begin{equation}\label{eq6_28_1}\begin{split}
F_{\text{Cas}}^{\parallel}=& \frac{ATN_D }{2^{p-2}\pi^{\frac{p-1}{2}} \Gamma\left( \frac{p-1}{2} \right)}
\sum_{l=-\infty}^{\infty}\sum_{\boldsymbol{n}\in\mathbb{Z}^q}\sum_{j=1}^{\infty}(-1)^j\int_{m_{\boldsymbol{n},l }}^{\infty}
y^2\left(y^2-m_{\boldsymbol{n},l}^2  \right)^{ \frac{p-3}{2}}
  e^{-2jay}dy\\
  =&  \frac{ATN_D }{2^{p-1}\pi^{\frac{p}{2}}  }
\sum_{l=-\infty}^{\infty}\sum_{\boldsymbol{n}\in\mathbb{Z}^q}\sum_{j=1}^{\infty}(-1)^j\left(\frac{m_{\boldsymbol{n},l}}{ja}\right)^{\frac{p}{2}}\left(2jam_{\boldsymbol{n},l}
K_{\frac{p+2}{2}}\left(2jam_{\boldsymbol{n},l}\right)-K_{\frac{p}{2}}\left(2jam_{\boldsymbol{n},l}\right)\right),
\end{split}\end{equation}where now $m_{\boldsymbol{n},l}=\sqrt{\lambda_{\boldsymbol{n}}^2+\xi_l^2}$; and
\begin{equation}\label{eq6_28_2}\begin{split}
F_{\text{Cas}}^{\parallel,T=0}=& \frac{AN_D }{2^{p}\pi^{\frac{p+1}{2}}  }
 \sum_{\boldsymbol{n}\in\mathbb{Z}^q}\sum_{j=1}^{\infty}(-1)^j\left(\frac{\lambda_{\boldsymbol{n}}}{ja}\right)^{\frac{p+1}{2}}\left(2ja\lambda_{\boldsymbol{n}}
K_{\frac{p+3}{2}}\left(2ja\lambda_{\boldsymbol{n}}\right)-K_{\frac{p+1}{2}}\left(2ja\lambda_{\boldsymbol{n}}\right)\right).
\end{split}\end{equation}
\eqref{eq6_28_1} and \eqref{eq6_28_2} are also the leading terms for the finite temperature Casimir force and the zero temperature Casimir force when $am\ll 1$. For the temperature correction, one put directly $m=0$ in \eqref{eq6_23_12} and uses \eqref{eq6_28_4} in the untwisted case.

Next, let us consider the small separation limit where $am\ll 1$ and $a\ll L_j$ for $j=p+1,\ldots,D$. In this case, we replace the summation over $\boldsymbol{n}\in \mathbb{Z}^q$ to integration over $\boldsymbol{n}\in \mathbb{R}^q$, which is equivalent to replacing $p$ in \eqref{eq6_28_1} by $D$. We find that for the finite temperature Casimir force,
\begin{equation}\label{eq6_28_5}\begin{split}
F_{\text{Cas}}^{\parallel}\sim & \frac{ATN_D }{2^{D-2}\pi^{\frac{D-1}{2}} \Gamma\left( \frac{D-1}{2} \right)}
\sum_{l=-\infty}^{\infty} \sum_{j=1}^{\infty}(-1)^j\int_{\xi_{l }}^{\infty}
y^2\left(y^2-\xi_{ l}^2  \right)^{ \frac{D-3}{2}}
  e^{-2jay}dy\\
  =& -\frac{ATN_D\Gamma(D) }{2^{2D-2}\pi^{\frac{D-1}{2}} \Gamma\left( \frac{D-1}{2} \right)a^D}(1-2^{1-D})\zeta_R(D)+ \frac{ATN_D }{2^{D-2}\pi^{\frac{D}{2}}  }
\sum_{l=1}^{\infty} \sum_{j=1}^{\infty}(-1)^j\left(\frac{\xi_{ l}}{ja}\right)^{\frac{p}{2}}\left(2ja\xi_{ l}
K_{\frac{D+2}{2}}\left(2ja\xi_{ l}\right)-K_{\frac{D}{2}}\left(2ja\xi_{ l}\right)\right).
\end{split}\end{equation}This can also be interpreted as the finite temperature Casimir force acting on a pair of parallel plates in $(D+1)$-dimensional Minkowski spacetime. In the high temperature  ($aT\gg 1$) limit, we find that
$$F_{\text{Cas}}^{\parallel}\sim -\frac{ATN_D\Gamma(D) }{2^{2D-2}\pi^{\frac{D-1}{2}} \Gamma\left( \frac{D-1}{2} \right)a^D}(1-2^{1-D})\zeta_R(D).$$In the low temperature
($aT\ll 1$) limit, \eqref{eq6_28_2} shows that the Casimir force is dominated by
$$F_{\text{Cas}}^{\parallel,T=0}\sim -\frac{A N_D\Gamma(D+1) }{2^{2D}\pi^{\frac{D}{2}} \Gamma\left( \frac{D}{2} \right)a^{D+1}}(1-2^{-D})\zeta_R(D+1).$$For the temperature correction, since in the limit $m=0$, $F(z)=z\cos az$, we deduce from \eqref{eq6_23_12} that
\begin{equation*}
\begin{split}
\Delta_TF_{\text{Cas}}^{\parallel}\sim &-\frac{A N_D}{\pi^{\frac{D+1}{2}}}\Gamma\left(\frac{D+1}{2}\right)(1-2^{-D})\zeta_R(D+1)T^{D+1}-\frac{AN_D \pi T^{\frac{D-2}{2}}}{2^{\frac{D-2}{2}} a^{\frac{D+4}{2}}}\sum_{l=1}^{\infty}(-1)^l
\sum_{k=0}^{\infty}\frac{\left(k+\frac{1}{2}\right)^{\frac{D+2}{2}}}{l^{\frac{D-2}{2}}}K_{\frac{D-2}{2}}\left(\frac{\pi\left(k+\frac{1}{2}\right)l}{aT}\right).
\end{split}
\end{equation*}Notice that the leading term is of order $T^{D+1}$.

\section{Conclusion}
In this article, we have investigated the finite temperature Casimir effect on a pair of parallel plates in a $(D+1)$-dimensional spacetime due to the vacuum fluctuations of a massive fermionic field with MIT bag boundary conditions on the plates. We assume that  $q<D$ of the dimensions are compactified to a torus and the field assumes general periodicity conditions on these compact dimensions. The Casimir free energy is computed using zeta function techniques and generalized Abel-Plana summation formula. Piston approach is employed to regularize the Casimir free energy. Low and high temperature expressions of the Casimir free energy and the Casimir force are derived. The asymptotic limits of the Casimir force when the separation between the plates is small are computed.

The most important result we obtain in this paper is that the   force acting on the plates is always attractive at any temperature and for any boundary conditions assumed on the compact dimensions. This extends the result obtained in \cite{7} for the zero temperature case. On the other hand, we also observe that this result is the same as the case of a scalar field with the same Robin condition on both plates \cite{6,1}. It can be considered as a manifestation of the principle that the Casimir force between two bodies with the same property is attractive \cite{2,3}.
\appendix
\section{Low temperature expansion of the Casimir free energy  }\label{a1}

Let $f(z)$ be a function that does not have poles on the positive real axis. Assume that $f(z)$ is    such that the functions $f_0(z), f_+(z)$ and $f_-(z)$ defined by
\begin{align*}
f_0(z) &= f(z)\frac{d}{dz}\ln\left(e^{i\pi z}+e^{-i\pi z}\right) = i\pi f(z)\left(1 - \frac{2}{e^{2i\pi z}+1}\right),\\
f_{\pm}(z) &= f(z)\frac{d}{dz}\ln\left(e^{\mp i\pi z}\right) = \mp i\pi f(z),
\end{align*}
satisfy \eqref{eq6_24_2}.
As in \cite{6}, one can derive the following formula from \eqref{eq6_24_1}:
\begin{equation}\label{eq6_23_6}
\begin{split}
 \sum_{l=0}^{\infty} f\left(l+\frac{1}{2}\right)
=& \int_0^{\infty}f(x)dx -i\int_0^{\infty}\frac{f(iy)-f(-iy)}{e^{2\pi y}+1}dy \\
& -\pi i\sum_{y>0} \frac{\text{Res}_{z=iy}f(z)-\text{Res}_{z=-iy}f(z)}{e^{2\pi y}+1}- 2\pi i\sum_{\substack{\text{Re}\; z> 0\\ \text{Im}\; z > 0}} \frac{\text{Res}_z f(z)}{e^{-2i\pi z}+1} + 2\pi i\sum_{\substack{\text{Re} z> 0\\ \text{Im}\; z < 0}} \frac{\text{Res}_z f(z)}{e^{2i\pi z}+1}.
\end{split}
\end{equation}
To derive the low temperature expansion for the Casimir free energy \eqref{eq6_23_1}, keep in mind that $m_{\boldsymbol{n},l}^2=m_{\boldsymbol{n}}^2+[2\pi(l+1/2)T]^2$ and notice that if $f$ is an even function,
\begin{equation*}
\sum_{l=-\infty}^{\infty}f\left(l+\frac{1}{2}\right)=2\sum_{l=0}^{\infty}f\left(l+\frac{1}{2}\right).
\end{equation*}Let
\begin{equation*}
\begin{split}
f(z;\boldsymbol{n})=&\frac{p-1}{2}\int_{\sqrt{m_{\boldsymbol{n}}^2+[2\pi T z]^2}}^{\infty}y\left(y^2-m_{\boldsymbol{n}}^2-[2\pi Tz]^2 \right)^{ \frac{p-3}{2}}
 \ln\left(1+\frac{y-m}{y+m}e^{-2ay}\right)dy
\\=&\int_0^{\infty}\frac{a-\frac{m}{y^2+m_{\boldsymbol{n}}^2+[2\pi Tz]^2-m^2}}{
\frac{\sqrt{y^2+m_{\boldsymbol{n}}^2+[2\pi T z]^2} +m}{\sqrt{y^2+m_{\boldsymbol{n}}^2+[2\pi T z]^2} -m}e^{2a\sqrt{y^2+m_{\boldsymbol{n}}^2+[2\pi T z]^2}}+1}\frac{y^pdy}{\sqrt{y^2+m_{\boldsymbol{n}}^2+[2\pi Tz]^2}}
=\int_0^{\infty}g(y,z;\boldsymbol{n})y^pdy,
\end{split}
\end{equation*}so that
\begin{equation}\label{eq6_23_7}
E_{\text{Cas}}^{\parallel}= -\frac{ATN_D }{2^{p-2}\pi^{\frac{p-1}{2}} \Gamma\left( \frac{p+1}{2} \right)}
\sum_{l=0}^{\infty}\sum_{\boldsymbol{n}\in\mathbb{Z}^q}f\left(l+\frac{1}{2};\boldsymbol{n}\right).
\end{equation}Then straightforward computation gives
\begin{equation*}
\begin{split}
\int_0^{\infty}f(x;\boldsymbol{n})dx=&\frac{p-1}{2}\frac{1}{2\pi T}\int_0^{\infty}\int_{\sqrt{m_{\boldsymbol{n}}^2+x^2}}^{\infty}y\left(y^2-m_{\boldsymbol{n}}^2-x^2 \right)^{ \frac{p-3}{2}}
 \ln\left(1+\frac{y-m}{y+m}e^{-2ay}\right)dydx\\
=&\frac{1}{4\sqrt{\pi}T}\frac{\Gamma\left(\frac{p+1}{2}\right)}{\Gamma\left(\frac{p}{2}\right)}\int_{m_{\boldsymbol{n}}}^{\infty}y\left(y^2- m_{\boldsymbol{n}}^2 \right)^{\frac{p-2}{2}}\ln\left(1+\frac{y-m}{y+m}e^{-2ay}\right)dy,\end{split}
\end{equation*}\begin{equation*}
\begin{split}
-i\int_0^{\infty}\frac{f(iu;\boldsymbol{n})-f(-iu;\boldsymbol{n})}{e^{2\pi u}+1}du=&-\frac{1}{2\pi T}\int_0^{\infty}\int_{\sqrt{y^2+m_{\boldsymbol{n}}^2}}^{\infty}\left(a-\frac{m}{y^2+m_{\boldsymbol{n}}^2-u^2-m^2}\right)
\frac{1}{e^{\frac{u}{T}}+1}\frac{du}{\sqrt{u^2-y^2-m_{\boldsymbol{n}}^2}}y^pdy\\
=&-\frac{1}{2\pi T}\int_{m_{\boldsymbol{n}}}^{\infty}\int_{0}^{\sqrt{u^2-m_{\boldsymbol{n}}^2}} \left(a+\frac{m}{y^2+ m^2}\right)
\left(u^2-m_{\boldsymbol{n}}^2-y^2\right)^{\frac{p-1}{2}}dy\frac{du}{e^{\frac{u}{T}}+1}\\
=&\frac{a}{4\sqrt{\pi}T}\frac{\Gamma\left(\frac{p+1}{2}\right)}{\Gamma\left(\frac{p+2}{2}\right)}\int_{m_{\boldsymbol{n}}}^{\infty}
\left(u^2-m_{\boldsymbol{n}}^2\right)^{\frac{p}{2}}\sum_{l=1}^{\infty}(-1)^le^{-\frac{lu}{T}}du\\&+
\frac{1}{2\pi T}\int_{0}^{\infty}\int_{\sqrt{y^2+m_{\boldsymbol{n}}^2}}^{\infty}  \frac{m}{y^2+ m^2}
\left(u^2-m_{\boldsymbol{n}}^2-y^2\right)^{\frac{p-1}{2}}\sum_{l=1}^{\infty}(-1)^le^{-\frac{lu}{T}}dudy\\
=&\frac{a\Gamma\left(\frac{p+1}{2}\right)}{4\pi T}\sum_{l=1}^{\infty}(-1)^l\left(\frac{2m_{\boldsymbol{n}}T}{l}\right)^{\frac{p+1}{2}}K_{\frac{p+1}{2}}\left(\frac{lm_{\boldsymbol{n}}}{T}\right)
\\&+\frac{\Gamma\left(\frac{p+1}{2}\right)}{2\pi^{\frac{3}{2}} T}\sum_{l=1}^{\infty}(-1)^l\int_0^{\infty} \frac{m}{y^2+ m^2} \left(\frac{2T\sqrt{y^2+m_{\boldsymbol{n}}^2}}{l}\right)^{\frac{p}{2}}K_{\frac{p}{2}}\left(\frac{l\sqrt{y^2+m_{\boldsymbol{n}}^2}}{T}\right)dy.
\end{split}
\end{equation*}
On the other hand, all the poles of $g(y,z;\boldsymbol{n})$ are on the imaginary axis, and they are given by
\begin{equation*}
z=\pm \frac{i}{2\pi T}\sqrt{k_1^2+y^2+m_{\boldsymbol{n}}^2},
\end{equation*}where $k_1$ are the solutions of $F(z)=0$. Therefore, the last two terms in \eqref{eq6_23_6} are zero, whereas
\begin{equation*}
\begin{split}
-\pi i\sum_{y>0} \frac{\text{Res}_{z=iy}f(z;\boldsymbol{n})-\text{Res}_{z=-iy}f(z;\boldsymbol{n})}{e^{2\pi y}+1}=&\frac{1}{2T} \sum_{k_1>0\,:\,F(k_1)=0}\int_0^{\infty}
\frac{1}{\exp\left(\frac{\sqrt{k_1^2+y^2+m_{\boldsymbol{n}}^2}}{T}\right)+1}\frac{y^pdy}{\sqrt{k_1^2+y^2+m_{\boldsymbol{n}}^2}}\\
=&\frac{1}{2T} \sum_{k_1>0\,:\,F(k_1)=0}\int_{\sqrt{k_1^2 +m_{\boldsymbol{n}}^2}}^{\infty}
\frac{\left(y^2-k_1^2-m_{\boldsymbol{n}}^2\right)^{\frac{p-1}{2}}}{\exp\left(\frac{y}{T}\right)+1}dy\\
=&-\frac{\Gamma\left(\frac{p+1}{2}\right)}{2\sqrt{\pi} T}\sum_{k_1>0\,:\,F(k_1)=0}\sum_{l=1}^{\infty}(-1)^l  \left(\frac{2T\sqrt{k_1^2+m_{\boldsymbol{n}}^2}}{l}\right)^{\frac{p}{2}}K_{\frac{p}{2}}\left(\frac{l\sqrt{k_1^2+m_{\boldsymbol{n}}^2}}{T}\right).
\end{split}
\end{equation*}Therefore,
\begin{equation*}\begin{split}
E_{\text{Cas}}^{\parallel}=& -\frac{AN_D }{2^{p}\pi^{\frac{p}{2}} \Gamma\left( \frac{p}{2} \right)}
\sum_{\boldsymbol{n}\in\mathbb{Z}^q}\int_{m_{\boldsymbol{n}}}^{\infty}y\left(y^2- m_{\boldsymbol{n}}^2 \right)^{\frac{p-2}{2}}\ln\left(1+\frac{y-m}{y+m}e^{-2ay}\right)dy\\&-\frac{AN_D }{2^{p-2}\pi^{\frac{p-1}{2}}}
\sum_{\boldsymbol{n}\in\mathbb{Z}^q}\sum_{l=1}^{\infty}(-1)^l
\Biggl\{\frac{a}{4\pi } \left(\frac{2m_{\boldsymbol{n}}T}{l}\right)^{\frac{p+1}{2}}K_{\frac{p+1}{2}}\left(\frac{lm_{\boldsymbol{n}}}{T}\right)
\\&\hspace{3cm}+\frac{1}{2\pi^{\frac{3}{2}} }\int_0^{\infty} \frac{m}{y^2+ m^2} \left(\frac{2T\sqrt{y^2+m_{\boldsymbol{n}}^2}}{l}\right)^{\frac{p}{2}}K_{\frac{p}{2}}\left(\frac{l\sqrt{y^2+m_{\boldsymbol{n}}^2}}{T}\right)dy\\
&\hspace{3cm}-\frac{1}{2\sqrt{\pi} }\sum_{k_1>0\,:\,F(k_1)=0}  \left(\frac{2T\sqrt{k_1^2+m_{\boldsymbol{n}}^2}}{l}\right)^{\frac{p}{2}}K_{\frac{p}{2}}\left(\frac{l\sqrt{k_1^2+m_{\boldsymbol{n}}^2}}{T}\right)\Biggr\}.
\end{split}\end{equation*}
\vspace{1cm}
\begin{acknowledgments}
 This project is funded by the Ministry of Higher Education of Malaysia   under FRGS grant FRGS/2/2010/SG/UNIM/02/2.
\end{acknowledgments}

\end{document}